\documentclass[a4paper,12pt, epsfig]{article}
\usepackage{epsfig}
\usepackage{amssymb}
\usepackage{amsfonts}

\newskip\humongous \humongous=0pt plus 1000pt minus 1000pt

\newif\ifdtup

\catcode`\@=11

\@addtoreset{equation}{section}
\def\theequation{\thesection.\arabic{equation}}

\def\@normalsize{\@setsize\normalsize{15pt}\xiipt\@xiipt
\abovedisplayskip 14pt plus3pt minus3pt%
\belowdisplayskip \abovedisplayskip
\abovedisplayshortskip \z@ plus3pt%
\belowdisplayshortskip 7pt plus3.5pt minus0pt}

\def\small{\@setsize\small{13.6pt}\xipt\@xipt
\abovedisplayskip 13pt plus3pt minus3pt%
\belowdisplayskip \abovedisplayskip
\abovedisplayshortskip \z@ plus3pt%
\belowdisplayshortskip 7pt plus3.5pt minus0pt
\def\@listi{\parsep 4.5pt plus 2pt minus 1pt
      \itemsep \parsep
      \topsep 9pt plus 3pt minus 3pt}}

\relax

\catcode`@=12

\catcode`\@=11

\def\section{\@startsection{section}{1}{\z@}{3.5ex plus 1ex minus
    .2ex}{2.3ex plus .2ex}{\large\bf}}

\def\thesection{\arabic{section}}
\def\thesubsection{\arabic{section}.\arabic{subsection}}

\def\appendix{\setcounter{section}{0}
  \def\thesection{Appendix \Alph{section}}
  \def\thesubsection{\Alph{section}.\arabic{subsection}}
  \def\theequation{\Alph{section}.\arabic{equation}}}

\def\SymBoxes#1#2#3#4{\newdimen\un@t \un@t#3%
\raisebox{#1}{\rule{#2\un@t}{#4}\hskip-#2\un@t
\@tempdimb\un@t \advance\@tempdimb by-#4\@tempcntb#2\relax%
\@whilenum{\@tempcntb>0}\do{
\rule{#4}{\un@t}\hskip\@tempdimb \advance\@tempcntb by\m@ne}%
\hskip-#2\un@t \rule[\un@t]{#2\un@t}{#4}%
\rule[\un@t]{#4}{#4}\hskip-#4
\rule{#4}{\un@t}}\hskip-#4}                

\begin{document}


\newcommand{\dd}{\textrm{d}}

\newcommand{\beq}{\begin{equation}}
\newcommand{\eeq}{\end{equation}}
\newcommand{\bea}{\begin{eqnarray}}
\newcommand{\eea}{\end{eqnarray}}
\newcommand{\beas}{\begin{eqnarray*}}
\newcommand{\eeas}{\end{eqnarray*}}
\newcommand{\defi}{\stackrel{\rm def}{=}}
\newcommand{\non}{\nonumber}
\newcommand{\bquo}{\begin{quote}}
\newcommand{\enqu}{\end{quote}}
\renewcommand{\(}{\begin{equation}}
\renewcommand{\)}{\end{equation}}
\def\de{\partial}
\def\Om{\ensuremath{\Omega}}
\def\Tr{ \hbox{\rm Tr}}
\def\H{ \hbox{\rm H}}
\def\HE{ \hbox{$\rm H^{even}$}}
\def\HO{ \hbox{$\rm H^{odd}$}}
\def\HEO{ \hbox{$\rm H^{even/odd}$}}
\def\HOE{ \hbox{$\rm H^{odd/even}$}}
\def\HHO{ \hbox{$\rm H_H^{odd}$}}
\def\HHEO{ \hbox{$\rm H_H^{even/odd}$}}
\def\HHOE{ \hbox{$\rm H_H^{odd/even}$}}
\def\K{ \hbox{\rm K}}
\def\Im{ \hbox{\rm Im}}
\def\Ker{ \hbox{\rm Ker}}
\def\const{\hbox {\rm const.}}
\def\o{\over}
\def\im{\hbox{\rm Im}}
\def\re{\hbox{\rm Re}}
\def\bra{\langle}\def\ket{\rangle}
\def\Arg{\hbox {\rm Arg}}
\def\exo{\hbox {\rm exp}}
\def\diag{\hbox{\rm diag}}
\def\longvert{{\rule[-2mm]{0.1mm}{7mm}}\,}
\def\a{\alpha}
\def\dag{{}^{\dagger}}
\def\tq{{\widetilde q}}
\def\p{{}^{\prime}}
\def\W{W}
\def\N{{\cal N}}
\def\hsp{,\hspace{.7cm}}
\def\bo{\ensuremath{\hat{b}_1}}
\def\bfo{\ensuremath{\hat{b}_4}}
\def\co{\ensuremath{\hat{c}_1}}
\def\cfo{\ensuremath{\hat{c}_4}}
\newcommand{\C}{\ensuremath{\mathbb C}}
\newcommand{\Z}{\ensuremath{\mathbb Z}}
\newcommand{\R}{\ensuremath{\mathbb R}}
\newcommand{\rp}{\ensuremath{\mathbb {RP}}}
\newcommand{\cp}{\ensuremath{\mathbb {CP}}}
\newcommand{\vac}{\ensuremath{|0\rangle}}
\newcommand{\vact}{\ensuremath{|00\rangle}                    }
\newcommand{\oc}{\ensuremath{\overline{c}}}

\newcommand{\Vol}{\textrm{Vol}}

\newcommand{\half}{\frac{1}{2}}
\begin{titlepage}
\begin{flushright}
SISSA 82/2008/EP
\end{flushright}
\bigskip
\def\thefootnote{\fnsymbol{footnote}}

\begin{center}
{\large {\bf
Contrasting confinement in superQCD and superconductors
  } }
\end{center}

\bigskip
\begin{center}
{\large  Jarah Evslin\footnote{\texttt{evslin@sissa.it}}}
\end{center}

\renewcommand{\thefootnote}{\arabic{footnote}}

\begin{center}
\vspace{1em}
{\em  { SISSA,\\
Via Beirut 2-4,\\
I-34014, Trieste, Italy\\
\vskip .4cm}}

\end{center}

\vspace{1.1cm}

\noindent
\begin{center} {\bf Abstract} \end{center}

\noindent
The vacuum of supersymmetric gauge theories (SQCD) with $\mathcal{N}=2$ softly broken to $\mathcal{N}=1$ resembles that of a BCS superconductor in that it has a condensate which collimates flux into vortices, leading to confinement.  We embed the SQCD vortex into the BCS theory by identifying the $\mathcal{N}=1$ vector multiplet mass and lightest massive chiral multiplet mass with the Fermi velocity divided by the London penetration depth and coherence length respectively.  Thus embedded the superconductivity is type I and so the vortex core is smaller than the coherence length.  Therefore nonlocal effects (Pippard electrodynamics) imply that the vortex solution is beyond the range of validity of the Landau-Ginzburg approximation implicit in the gauge theory.  In other words, the vortex solution contains gradients greater than those for which the BCS and gauge theory descriptions agree.  We consider more general superpotentials which are polynomial in the chiral multiplets and find that, unless one adds a SUSY breaking sector, one obtains type II superconductivity only when the superpotential perturbation is at least quadratic in the fundamental chiral multiplets and at least linear in the adjoint chiral multiplets, in which case there is no $\mathcal{N}=2$ supersymmetry in the ultraviolet.


\vfill

\begin{flushleft}
{\today}
\end{flushleft}
\end{titlepage}

\hfill{}


\setcounter{footnote}{0}

\section{Introduction}

While the Lagrangian description of QCD has been known since before I was born, its ground state remains a mystery.  For example, it is not known whether it is translation-invariant \cite{NO}.  Experiments and lattice calculations confirm that colored objects are confined.  This confinement appears to be caused by vortices which repel at large distances.  Yet none of this has been demonstrated analytically.  

Perhaps the largest breakthrough in this direction is Seiberg and Witten's solution of $\mathcal{N}=2$ super QCD \cite{SW,SW2}.  Softly breaking the supersymmetry to $\mathcal{N}=1$ one can analytically find the condensate and its vortices, thus demonstrating confinement.  The critical question then becomes, just how similar are these vortices to those of our world's QCD?

The first clue that they differ is that vortices in ordinary Yang-Mills, like those in a type II superconductor, repel \cite{hep-lat/0607014}.  Vortices in SQCD classically break the gauge symmetry and so come with orientations, although in some quantum theories the symmetry is restored by instantons \cite{noi6}.  In this note we will restrict our attention to classical vortices.  Intuitively there are two forces between vortices, an attractive force mediated by scalars and a repulsive force mediated by gluons.  The least massive particle exerts the strongest force at long distances.  In the case of theories with BPS vortices, for example those with a FI term and no superpotential, the scalars and gluons have the same mass and so the vortices neither attract nor repel.  When the supersymmetry is broken by a superpotential polynomial in the adjoint chiral multiplets, then the vortices attract at long distances.  In either case, they do not behave like the vortices in Yang-Mills or in a superconductor.

To make the analogy between supersymmetric gauge theories and superconductors more precise, we will provide an identification between the parameters of the gauge theory and the parameters of the low energy Landau-Ginzburg effective theory of the superconductor.  With this identification, one may ask whether a given supersymmetric gauge theory corresponds to a type I or a type II superconductor.  Our main result is that, in the absence of an additional supersymmetry breaking sector, softly broken $\mathcal{N}=2$ supersymmetric QCD leads to type I superconductivity and so attractive vortices.  Conversely we find some supersymmetric gauge theories with $\mathcal{N}=1$ supersymmetry at all energy scales which are always identified with type II superconductors and so possess repulsive vortices.

We will begin in Sec.~\ref{ubersec} with a review of the BCS theory of superconductivity and its low energy effective description.  In particular we will describe the two characteristic length scales, the London penetration depth and the coherence length, and we will describe their relations to vortex solutions and the critical magnetic field.  Next in Sec.~\ref{sqcdsec} we will describe $\mathcal{N}=2$ supersymmetric QCD softly broken to $\mathcal{N}=1$, and motivate the identification of the inverse masses of its supermultiplets with the length scales of the superconductor.  In Sec.~\ref{cartoonsec} we will describe the possibility of adding superpotential terms to the gauge theory which are polynomial in the adjoint chiral multiplets, and will argue that no such term yields type II superconductivity, but some may lead to repulsive vortices.  Finally in Sec.~\ref{gensec} we consider superpotentials which depend on both the adjoint and the fundamental chiral multiplets and find that some do lead to type II superconductivity, but that in these cases the $\mathcal{N}=2$ supersymmetry is not restored in the ultraviolet.

\section{Length scales in the BCS theory of superconductivity} \label{ubersec}

Low temperature superconductors are in general well described by a microscopic theory called the BCS theory of superconductivity.  When gradients are sufficiently small and the temperature is close to the critical temperature, this theory possesses a macroscopic limit, a low energy effective theory called the Landau-Ginzburg theory.  This is just a U(1) gauge theory with a charged complex scalar in a Mexican hat potential.  We wish to compare the low energy effective theory with supersymmetric QCD.  However the effective theory only approximates a Landau-Ginzburg theory when gradients are sufficiently small, otherwise one must consider nonlocal corrections which arise from the correlations of macroscopically separated electrons in the BCS theory.  Theories with such nonlocal terms are said to exhibit Pippard electrodynamics.  This nonlocality has no analogue in the supersymmetric gauge theory, and so the superconductor-SQCD analogy breaks down at the length scale at which they become relevant.

Superconductors have two characteristic scales.  The first is called the London penetration depth $\lambda$.  It is the characteristic distance over which the magnetic field decays.  In our case it will determine the characteristic width of a magnetic vortex.  Except for semilocal vortices, which appear to be unstable in the non-BPS case \cite{0810.5679}, as we will review below the vortices naturally appearing in SQCD have a magnetic field which decays exponentially at large radius
\beq
A\sim e^{-r/\lambda}.
\eeq
The coefficient of this decay may be identified with the penetration depth.

The other characteristic scale is the coherence length $\xi$.  Intuitively this is the minimum size of a Cooper pair wavepacket allowed by the uncertainty principle, and so it sets the scale of the nonlocality mentioned above.  In the Landau-Ginzburg theory a Cooper pair is treated as a single quasiparticle.  Its energy $E$ is equal to the band gap between the Cooper pair's actual energy and the energy that the electrons would have if they disassociated and moved into the conducting band.  The temporal uncertainty is just the reciprocal of the energy.  To obtain the spatial uncertainty, one need only multiply by its velocity, which is the Fermi velocity $v_F$, to obtain the coherence length
\beq
\xi=\frac{2\hbar v_F}{\pi E}. \label{xi}
\eeq

As usual the uncertainty principle reflects the fact that any attempt to trap a particle in a box results in a minimum energy.  In particular, if the box size is equal to or smaller then $\xi$ then the minimum energy will be greater than or equal to the band gap $E$.  If one increases the energy of a Cooper pair by the band gap then the electrons will have enough energy to escape into the conductance band, and the Cooper pair will disassociate.  Therefore if a vortex is smaller than $\xi$ then the energy will be sufficiently high that the Cooper pairs disassociate, and the core of the vortex will cease to be superconducting.  This does not imply that the magnetic flux turns off, indeed this may not be possible with given boundary conditions, but rather that it exceeds the critical magnetic field and so the Cooper pair condensate field is no longer a good degree of freedom in its core.  

This means that when the vortex core is smaller than the coherence length, in general the paired electrons inside of the core will be paired with electrons outside of the core.  This means that the nonlocal effects caused by the coherence length are important at the scale of such vortices, and the Landau-Ginzburg approximation is invalid.  $\lambda$ may be less than, equal to, or greater than $\xi$.  In the first case the superconductor is said to be type I, in the third case type II.  We will refer to the second case as critical.  Recall that the magnetic field in a vortex in the BCS condensate has a characteristic size $\lambda$, and so for a type I superconductor $\lambda$ is less than $\xi$ and vortices will be beyond the range of the Landau-Ginzburg approximation.   


Conversely, in type II superconductors there may be vortices.   In fact the structure of a type II superconductor is a bit more complicated.  While type I superconductors are often simply elemental metals, type II superconductors are made of composites.  In particular in the presence of a magnetic field they arrange into a state known as the Abrikosov lattice, where the magnetic field penetrates in a regular lattice of vortices.  It would be interesting to see if a better understanding of vortices in this lattice could shed light on confinement in the color glass condensate model of the QCD vacuum, which has obtained considerable support at RHIC \cite{cgc}.  In particular, the stability of the Abrikosov lattice comes from the fact that the domain wall between the normal and superconducting phases has negative tension for type II superconductors, leading to a boundary that tries to maximize its area and making it energetically favorable to create pockets of nonenergetically favorable normal phase below the critical temperature.  One may search for such negative tension domain walls between two phases of ordinary lattice QCD.  In the presence of such walls, the lowest energy state of QCD may well be a nontranslationally-invariant configuration, like the Abrikosov-lattice, in which regions of phases which do not minimize energy exist due to the energy benefits of the negative tension domain walls, such as the pockets of free gluons in the color glass condensate.


\section{Vortices in SQCD} \label{sqcdsec}

We are interested in classical vortices in SQCD theories in which the supersymmetry is softly broken from $\mathcal{N}=2$ to $\mathcal{N}=1$, as constructed for example in Refs.~\cite{HT,noi6}.  Classically these vortices break the gauge group down to a product of abelian groups and a nonabelian group under which it is not charged.  Therefore classically these vortices are essentially abelian.  

For example, consider vortices in an SU(3) gauge theory broken to U(2), with $\mathcal{N}=2$ broken to $\mathcal{N}=1$ by a mass for the adjoint chiral multiplet.  Furthermore consider 5 fundamental flavors with bare masses chosen such that the U(2) charged hypermultiplets are massless.  Then each classical vortex solution breaks the $SU(2)$ quotient of the $U(2)$ to $U(1)$.  The possible breakings are parametrized by a 2-sphere $S^2$, and so the classical moduli space is $S^2$.  In fact, the low energy effective theory of the vortex is just a sigma model with target space $S^2$.  The interaction of a pair of vortices depends on their relative orientations in this moduli space.  However, using the fact that the unrenormalized couplings of the $U(1)$ and $SU(2)$ are equal, the contribution of the nonabelian degrees of freedom will never lead to a change in sign of the force between the vortices, although vortices with opposite orientations are decoupled \cite{0709.1910}.

Quantum mechanically instantons generate a potential on this $S^2$, as can been seen for example in the mirror dual Sine (sinh) Gordon theory.  This theory contains only two vacua, and the $U(2)$ gauge symmetry is restored.  The dynamics of the vortices in these quantum vacua has not yet been analyzed, and it would be interesting to see if they differ from that of the classical abelian vortices.  However, as the quantum vacuum is a superposition of abelian vacua, this would be surprising.

Therefore we will turn our attention to an entirely abelian theory, softly broken $\mathcal{N}=2$ SQED.  For simplicity, we first consider a breaking to $\N=1$ via a superpotential which is quadratic in the adjoint chiral superfield, or more generally a superpotential whose critical points are of degree at most equal to two.  The linear term may be interpreted as an FI F-term $\xi$, in the absence of higher order terms it is related to an FI D-term by the $\mathcal{N}=2$ R-symmetry.  The second derivative at the VEV of the scalar in the vector multiplet may be interpreted as a mass $\mu$ for the adjoint chiral multiplet $\Phi$.

Generically a squark condensate forms which breaks the gauge symmetry.  In Refs.~\cite{0012250,0303047,0709.1910} the authors found the mass spectrum of the excitations about the condensate vacuum.  As the theory preserves $\mathcal{N}=1$ supersymmetry, one need only determine the difference between the vector multiplet mass and the masses of the various chiral multiplets.  They began with the bosonic Lagrangian
\bea
L_b=&&\frac{1}{4e^2}F^2_{\mu\nu}+\frac{1}{e^2}|\partial_\mu\phi|^2+\overline{\nabla}_\mu\overline{q}\nabla_\mu q+\overline{\nabla}_\mu\overline{\tilde{q}}\nabla_\mu \tilde{q}\nonumber\\
&+&\frac{e^2}{8}(|q|^2-|\tilde{q}|^2)^2+\frac{e^2}{2}|\tilde{q}q+i\xi+\sqrt{2}\mu\phi|^2+\frac{1}{2}(|q|^2+|\tilde{q}|^2)|\phi|^2
\eea
where $\phi$ is the scalar in the vector multiplet and $e$ is the electric coupling constant.  The covariant derivatives are defined by
\beq
\nabla_\mu=\partial_\mu-\frac{i}{2}A_\mu\hsp
\tilde{\nabla}_\mu=\partial_\mu+\frac{i}{2}A_\mu.
\eeq
The linear term $\xi$ leads to a vacuum expectation value for $q$ and $\tilde{q}$
\beq
\langle q\rangle=i\langle \tilde{q}\rangle=\sqrt{\xi}
\eeq
but not for $\phi$.  

One may then expand all of the fields into classical and quantum contributions
\beq
q=\sqrt{\xi}+\delta q\hsp \tilde{q}=-i\sqrt{\xi}+\delta\tilde{q}\hsp \phi=\delta\tilde{\phi}.
\eeq
The mass matrix is obtained by taking the second derivatives of $V$ with respect to all of the quantum fields.  In particular, the mass for the $\mathcal{N}=1$ vector multiplet is given by the $\xi$ term, as it led to the squark VEV which breaks the gauge symmetry
\beq
m^2_{vec}=\frac{e^2\xi}{2}.
\eeq
The two uneaten chiral multiplets have a nondiagonal mass matrix whose eigenvalues are
\beq
m^2_{chi}=m^2_{vec}\left(1+\frac{e\mu}{\xi}\pm\sqrt{\frac{e^2\mu^2}{\xi^2}+\frac{4e\mu}{\xi}}\right). \label{mass1}
\eeq
The important feature here is that, for the eigenvalue with the minus sign, the factor in parentheses is less than one if $\mu\neq 0$ and is otherwise equal to one.  When $\mu=0$, the $\mathcal{N}=2$ symmetry is unbroken and the vortex is BPS.  Therefore we have learned that the least massive particles are always in the chiral multiplet, although in the BPS case they all have the same mass.

This fact is significant because the exchange of a chiral multiplet leads to an attractive force between vortices whereas whereas the exchange of a vector multiplet may lead to an attraction or repulsion depending on the relative orientations \cite{0202172}.  At long distances, the interaction is dominated by the lightest species of carrier, as was demonstrated in the nonabelian case in Ref.~\cite{0709.1910} where the authors found that in the nonabelian case the lightest multiplet is again a chiral multiplet.  Therefore at long distances SQCD vortices with the same orientation will in general attract, or in the BPS case will not exert a force.

The relation between the masses and the forces leads to a natural identification of the SQCD masses and the BCS theory length scales.

\noindent
{\bf{Proposed identification: }}{\it{The London penetration depth and coherence length of the BCS theory are the inverses of the vector multiplet and lightest massive chiral multiplet masses of the SQCD vacuum multiplied by the Fermi velocity:}}
\beq
\lambda=\frac{2\hbar v_F}{\pi m_{vec}}\hsp 
\xi=\frac{2\hbar v_F}{\pi m_{chi}}. \label{claim}
\eeq

In particular we recover that in the presence of an adjoint chiral multiplet mass $\mu$, the corresponding BCS theory satisfies $m_{chi}<m_{vec}$ and so $\lambda<\xi$, identifying it as a type I superconductor, as was found in Ref.~\cite{0709.1910}.  On the other hand, when $\mu$ goes to zero the theory becomes critical, as was found in Refs.~\cite{GF,0012250}.  The fact that vortices neither attract nor repel in the critical theory was demonstrated in Ref.~\cite{JR}.

The identification of $\lambda$ with the inverse photon mass follows from the fact that in the vortex solution the photon decays asymptotically like $e^{-m_{vec} r}$, which solves the Klein-Gordon equation with mass $\lambda$ and speed of light $v_F$.  

The identification of $\xi$ with the inverse chiral multiplet mass on the other hand, using Eq.~(\ref{xi}), is an identification of the squark mass with the energy gap of the Cooper pair, which is the amount of energy that it costs to disassociate the pair.  This generalizes the usual identification of the Higgs mass with the inverse coherence length in the nonsupersymmetric case.  In the gauge theory, which is relativistic, only the mass squared enters and so the sign of the mass is unimportant.  Therefore this energy is identified with the energy that one requires to create a squark from the vacuum, which is natural as the squark condensate is replaced with the Cooper pair condensate.

In the identification one considers only the massive chiral multiplets because the massless multiplet is eaten by the Higgs mechanism.  In section~\ref{gensec} we will consider SUSY gauge theories with an additional superpotential term which leads to type II superconductivity.  In this case the lightest massive real scalar is part of the massive $\mathcal{N}=1$ vector multiplet, and it is the next lightest multiplet which must be used in our identification.

With the identification (\ref{claim}) we have concluded that the BCS theory corresponding to our SQCD, without the hard breaking considered in section~\ref{gensec}, is either type I or marginal.  This is in stark contrast to ordinary Yang-Mills.  Here lattice simulations indicate that, as a superconductor, the Yang-Mills vacuum is just barely type II \cite{hep-lat/0607014}.  This means that there is a slight repulsive force between vortices.  These lattice results may also be consistent with marginal superconductivity and a higher order interaction.

\section{Superpotentials that are polynomial in the adjoint matter} \label{cartoonsec}


In the rest of this note we will try to understand whether a general superpotential would allow for type II vortices.  For example, one may consider a superpotential polynomial in the adjoint chiral multiplet with a critical point of degree at least three, so that there is an FI term and an interaction term but no mass term. In this section we will argue that higher degree superpotentials constructed entirely from adjoint fields do not lead to type II superconductivity, but may lead to vortices which repel and so may be consistent with lattice simulations of Yang-Mills.  In fact, such a repulsion in a critical superconductor would be phenomenologically similar to the slightly type II superconductivity observed on the lattice.

\begin{figure}
\begin{center}
\includegraphics[width=0.6\textwidth]{hw.eps}
\caption{A $\mathcal{N}=2$ theory broken to $\mathcal{N}=1$ by a superpotential that is polynomial in the adjoint chiral multiplet can be engineered in type IIA string theory.  There are two NS5-branes, one which is flat, and the other whose embedding is described by the derivative of the superpotential.  The gauge theory lives on a semi-infinite D4-brane, describing a locked flavor and color, which extends from the flat NS5-brane to infinity.  The vortices are D2-branes extending from the D4-brane to the curved NS5-brane.  The length of the D2 yields the vortex tension.}
\label{hwfig}
\end{center}
\end{figure}

A Hanany-Witten brane cartoon realization gives some hope that cubic superpotentials may lead to qualitatively new behavior.  We review the representation of a vortex in type IIA string theory presented in Ref.~\cite{ABE}.  Consider 10-dimensional Minkowski space.  There are 2 NS5-branes, both extended in the directions $x^{0}$ through $x^5$.  One is flat and is located at $x^{6}$ through $x^9$ all equal to zero.  We will refer to the other as the curved NS5-brane, although it will not always be curved.  Its coordinate in the $x^6$ direction is $1/e^2$, and its $x^7$ coordinate is the FI parameter.  The superpotential $W$ will determine its location in the $x^8$ and $x^9$ directions.  We will consider a general polynomial superpotential, leading to the brane realization of Ref.~\cite{9708044}.  Intuitively
\beq
x^8+ix^9=W\p(x^4+ix^5). \label{sup}
\eeq
This may be seen in Fig.~\ref{hwfig}.  More rigorously, this is only true asymptotically at large $x^6$, corresponding to the weak coupling regime in the ultraviolet.  The gauge theory lives on a D4-brane which is stretched in the $x^6$ direction from the flat NS5-brane, past the other, to $x^6=+\infty$, and also is flat in the directions $x^0$ through $x^3$.  It is at the origin in all other directions.  

Notice that the D4-brane does not in general touch the curved NS5-brane, since it is always located at $x^4+ix^5=x^8+ix^9=0$.  Instead, in the approximation of Eq.~(\ref{sup}) turning off the FI term and taking the higher derivatives of the superpotential to be small, the closest approach is at a distance of $|W\p(0)|$ on the $x^8+ix^9$ plane.  A vortex in the field theory corresponds to a D2-brane which extends along $x^0$ and $x^3$ and also stretches between the D4-brane and the curved NS5-brane at this closest approach.  Its mass is therefore equal to the smallest distance between the D4 and NS5, at least when it is BPS.

For example, one may consider a theory with an FI D-term $\xi$ and a superpotential $W=a\phi$ by noting that, using the $SU(2)_R$ R-symmetry of $\mathcal{N}=2$, one may rotate to a configuration with no FI term and a real superpotential
\beq
W=\left(\sqrt{\xi^2+|a|^2}\right)\phi. \label{ten}
\eeq
As the NS5-brane is straight in this case, the minimum length D2-brane extends in the $x^8$ direction.  It therefore corresponds to a BPS vortex whose tension is equal to $\sqrt{\xi^2+|a|^2}$.  This is the correct value of the tension as calculated in the field theory \cite{ABE}.

What happens if we consider a quadratic superpotential, adding a mass term for the adjoint chiral multiplet?  Now the NS5-brane will be at an angle, as $x^4+ix^5$ will be proportional to $x^8+ix^9$.  As a result if the D2-brane vortex extends purely along the $x^8+ix^9$ direction then it will not minimize its length, as it will not be orthogonal to the NS5.  To minimize its length it needs to be orthogonal to the NS5-brane, and so in addition to extending along $x^0$ and $x^3$, it also will extend along a superposition of a direction in $x^4+ix^5$ and a direction in $x^8+ix^9$.  The curved NS5-brane is not curved in this case, but it is at an angle with respect to the other NS5-brane, which breaks the supersymmetry from $\mathcal{N}=2$ to $\mathcal{N}=1$.  The D2-brane, corresponding to the vortex, does not preserve the remaining supersymmetry and so is not BPS.  These are the vortices that we have been discussing.  They attract, unlike vortices in superconductors and in Yang-Mills.

The crucial question is whether the behavior is more or less the same for higher degree superpotentials.  We will now provide some mild hope that new phenomena occur in the cubic case.  This case, along with the linear and quadratic cases discussed above, may be seen in Fig.~\ref{essempifig}.  Imagine a superpotential with only a linear and a cubic term
\beq
W=a\phi+b\phi^3.
\eeq
For concreteness, let $a$ and $b$ be real.  Now a D2-brane may again extend in the $x^8$ direction from the D4-brane to the NS5.  It would have a length $a$, corresponding to a vortex of tension $a$.  Unlike the quadratic case, the intersection with the NS5-brane is orthogonal.  

This orthogonality is generic to superpotentials whose second derivative vanishes at the end of the vortex.  The vanishing second derivative implies that there is no additional contribution to the chiral multiplet mass matrix, and so the chiral multiplet and vector multiplet remain degenerate and the BCS theory remains critical.  However there will be higher order interactions, which we will now discuss.

\begin{figure}
\begin{center}
\includegraphics[width=0.7\textwidth]{essempi.eps}
\caption{Brane cartoons corresponding to linear, quadratic and cubic real superpotentials are shown.  In the linear case, both NS5-branes are parallel, the the theory is $\mathcal{N}=2$ supersymmetric and the vortex is BPS.  The D2-brane extends along the $x^8$ direction.  In the quadratic case supersymmetry is broken and the vortex extends in a diagonal direction.  In the cubic case the vortex may appear BPS at tree level if the curvature is sufficiently small, as the vortex again only extends in the $x^8$ direction.  The vortex worldvolume theory has a tachyon if the D2-brane length is not minimized by this configuration, in which case there is spontaneous symmetry breaking leading to the non-BPS configuration on the bottom right.}
\label{essempifig}
\end{center}
\end{figure}

The second derivative of the NS5-brane position is nonzero at the intersection, it is proportional to $b$.  Therefore, while the vortex extremizes its length in the $(x^4,x^8)$ plane, it does not necessarily minimize it.  In fact, the length will be minimized when
\beq
6ab\geq -1.
\eeq
Thus in this range one may expect stable vortices which nonetheless will not have the standard ANO radial profile, due to the additional interactions.   Therefore one may hope for qualitatively different interactions between such vortices.  On the other hand, this transition may simply be an artifact of the inapplicability of the approximation (\ref{sup}).  However this approximation works at $b=0$, reproducing (\ref{ten}), and so for sufficiently small $b$ perhaps it may still be trusted.

More concretely, the length of the D2-brane depends quadratically on its angle in the $(x^4,x^8)$ plane.  This angle will therefore be a massive mode on the vortex worldvolume, with a mass squared proportional to $6ab$.  In particular, when the mass squared is negative the mode will be tachyonic.  The D2-brane will then move to minimize its length, spontaneously breaking the $\Z_2$ symmetry which changes the sign of $\phi$, similarly to the spontaneous supersymmetry breaking of a similar model in Refs.~\cite{SY1,SY2}.  It will not decay, as it carries a conserved topological charge \cite{Stefano}, instead it will simply rotate until its length is minimized.  At the new intersection point of the D2-brane and the curved NS5-brane, the NS5-brane will be at an angle, and so the configuration will be a perturbation of the configuration corresponding to a slight adjoint chiral multiplet mass.  Therefore in this vacuum, as in the case of a quadratic superpotential, the chiral multiplet masses will split.  One mass will become lower than that of the vector multiplet, and one higher.  Therefore the superconductor will be type I.

On the other hand the region $6ab>-1$ is more promising.  The rotational mode is massive and so the spectrum appears to be tachyon free.  More generally this is the case with higher order superpotentials, as can be seen using the above brane cartoons or alternately from the field theory analysis of the similar model in Ref.~\cite{SY1}.  However this superpotential still does not contribute to the chiral multiplet mass matrix, and so again the London penetration depth and coherence lengths will be equal, leading to critical superconductivity and so marginally stable Cooper pairs.  The hope is then that the additional interactions may stabilize the Cooper pairs.  A simpler question is whether the higher order interactions lead to a repulsion between the vortices as in lattice Yang-Mills.

Consider a perturbation to the superpotential of the form
\beq
\Delta W=\epsilon \Phi^k
\eeq
where $k>2$.  Again this introduces no new quadratic terms in the potential and so does not contribute to the tree level mass matrices except, as in the cubic case, when it leads to spontaneous symmetry breaking and a new vacuum with a value of the adjoint scalar condensate at which the second derivative of the superpotential is nonzero.  But in this case again one chiral multiplet mass decreases while the other increases, and so the minimum mass multiplet is chiral and the superconductor is type I.

Far from the vortex, where the squark field satisfies the vacuum equations of motion, before one included the perturbation it fell off like $e^{-m_{chi}r}$, which solved the Klein-Gordon equation with mass $m_{chi}$.  The additional superpotential term modifies the Klein-Gordon equation to roughly
\beq
\partial^2 \phi = m_{chi}^2\phi+\epsilon k^2 \phi^{2k-3}.
\eeq
The perturbed solution now contains an extra piece
\beq
\phi\sim e^{-m_{chi}r} + \frac{\epsilon k^2}{((2k-3)^2-1)m_{chi}^2} e^{-(2k-3)m_{chi}r}.
\eeq
Notice that the perturbed solution always has a bigger condensate field then an unperturbed solution.  This is similar to the addition of an adjoint chiral mass, which shifted the exponent from $m_{vec}$ to $m_{chi}$ and so increased the amount of condensate at large radius.  As the condensate is responsible for an attractive force, this increase always leads to a net attraction.

The question now becomes whether the perturbed solution with the perturbed energy, when placed near another such vortex, leads to a higher or lower energy than the sums of the original energies.  At order $\epsilon$ there are two contributions to the interaction energy.  First there is the energy of the unperturbed solution with the perturbed potential $\epsilon \phi^{2k-2}$.  Then there is the energy of the perturbed solution of one vortex, with the unperturbed part of the other using the unperturbed quadratic potential.  This contribution is again of order $\epsilon \phi^{2k-2}$, and as was described above is attractive.  Therefore it is conceivable that for some choice of models, taking into account for the effect on the vector multiplet as well, the order $\epsilon$ contribution of the original solution will lead to more repulsion than the attraction caused by the unperturbed potential of the perturbed solution, and so the vortices will repel at large distance.  If such a superpotential can be found, then it is plausible that the same potential may also stabilize the Cooper pairs and so allow for a vacuum structure similar to that of the BCS theory.

\section{General polynomial superpotentials} \label{gensec}

We have argued that, while superpotentials which are higher order polynomials in the adjoint chiral multiplets may lead to repulsive vortices, they do not lead to tree level chiral multiplet masses which are all higher than the vector multiplet mass and so do not lead to type II superconductivity and BCS vortices.  As has been stressed by A. Adams, this is analogous to the fact that at least one selectron mass eigenvalue is, in contrast with experimental observation, lower than the tree level electron mass in simple models that break supersymmetry at tree level \cite{primer}.  In that case, there are two solutions to the problem.  First, one may consider a model in which the symmetry breaking or at least the mass splitting is dynamical.  Second, one may add additional material and break the symmetry in an additional sector.

In this section we will consider a third possibility which has no analogue in the selectron problem.  We consider superpotentials which are not only polynomial in the adjoint chiral multiplets $\Phi$, but also in the fundamental chiral multiplets $Q$ and $\tilde{Q}$.  In an $\mathcal{N}=1$ theory it is easy to obtain type II superconductivity, as one may fix the values of the masses of the chiral multiplets at will.  Therefore we will eventually be interested in theories in which the $\mathcal{N}=2$ supersymmetry is broken softly, and so in particular we will not be interested in superpotentials with 4 or more fundamental chiral multiplets.  This leaves us with a general superpotential of the form
\beq
\mathcal{W}=Q\Phi\tilde{Q}+F(\Phi)+QG(\Phi)\tilde{Q} \label{w}
\eeq
where $F$ and $G$ are functions of $\Phi$.  Some mild restrictions on these functions are placed by the absence of poles in the following calculation.  Note that for general functions $F$ and $G$ the $\mathcal{N}=2$ supersymmetry breaking will not be soft.

The superpotential (\ref{w}) together with the D term leads to the scalar potential energy
\bea
&&V=V_1+V_2+V_3\hsp V_1=\frac{g^2}{2}(|q|^2-|\tilde{q}^2|)\label{pot}\\
&&V_2=2g^2|\tilde{q}q+F\p+\tilde{q}qG\p|^2\hsp
V_3=2|\phi+G(\phi)|^2(|q|^2+|\tilde{q}|^2)\nonumber
\eea
where lower case letters are the scalar components of the chiral multiplets and $g$ is the coupling constant.  This potential energy is minimized by classical expectation values $\phi_0$, $q_0$ and $\tilde{q}_0$ which satisfy
\beq
\phi_0=-G(\phi_0)\hsp q_0=-\tilde{q}_0=\sqrt{\frac{F\p(\phi_0)}{1+G\p(\phi_0)}}. \label{min}
\eeq
In particular in the previous section $G$ is equal to zero and so the adjoint scalar VEV vanishes.  If $G$ is constant, corresponding to a bare squark mass, it may be absorbed into a shift in $\phi$ and so we will not be surprised to find that the mass spectrum is independent of the constant part of $G$.  In particular, if such a deformation leads to type II superconductivity then it will be at least the linear part of $G$ which contributes, which is a marginal operator and so ruins the $\mathcal{N}=2$ supersymmetry in the ultraviolet.

Expanding the fields about their minima (\ref{min})
\beq
\phi=-G(\phi_0)+g(a+ib)\hsp
q=\sqrt{\frac{F\p}{1+G\p}}+c+id\hsp
\tilde{q}=-\sqrt{\frac{F\p}{1+G\p}}+e+if
\eeq
we find that the terms in the potential energy (\ref{pot}) which contribute to the masses are
\bea
V_1&\supset&2g^2|\frac{F\p}{1+G\p}|[(c+e)\cos\phi+(d+f)\sin\phi]^2\nonumber\\
V_2&\supset&2g^2|F\p+F\p G\p|([e-c+gr(a\cos\theta+b\sin\theta)]^2+[f-d+gr(a\sin\theta-b\cos\theta)]^2)\nonumber\\
V_3&\supset&4g^2|F\p+F\p G\p|(a^2+b^2)
\eea
where the functions $F$ and $G$ and their derivatives are evaluated at the minima (\ref{min}) and we have defined
\beq
\phi=\frac{1}{2}\rm{Arg}\left(\frac{F\p}{1+G\p}\right)\hsp
\frac{\partial_\phi(F\p/(1+G\p))}{F\p/(1+G\p)}=re^{i\theta}.\label{param}
\eeq
We will see that the cut in the definition of $\phi$ is irrelevant, as the mass matrix eigenvalues are independent of $\phi$.

The vector multiplet mass is given by the Higgsing due to the squark VEV, as we are working in an abelian theory where the scalars $\phi$ are uncharged.  In the nonabelian theory the $\phi$ VEVs will contribute to the W-boson masses.  We will now see that the chiral multiplet masses depend only on the vector multiplet mass, on the parameter $r$ from Eq.~(\ref{param}), which roughly corresponds to the ratio of the adjoint chiral multiplet mass to the linear term in the previous subsection, and also on the parameter
\beq
x=|1+G\p|^2
\eeq
which is roughly the square of the new coefficient of the $Q\Phi\tilde{Q}$ term in the superpotential.  In particular, $x$ corresponds to a marginal deformation and any value different from $x=1$ implies that $\mathcal{N}=2$ SUSY remains broken in the ultraviolet.

The vector multiplet mass squared is then
\beq
m^2_\gamma=2g|\frac{F\p}{1+G\p}|.
\eeq
In the basis $(c,e,d,f,a,b)$ the potentials may be summarized by the chiral multiplet mass matrix\beq
M^2=\frac{m^2_\gamma}{2}\left(\begin{array}{cccccc}
\cos^2\phi+x&\cos^2\phi-x&\cos\phi\sin\phi&\cos\phi\sin\phi&xr\cos\theta&xr\sin\theta\\
\cos^2\phi-x&\cos^2\phi+x&\cos\phi\sin\phi&\cos\phi\sin\phi&-xr\cos\theta&-xr\sin\theta\\
\cos\phi\sin\phi&\cos\phi\sin\phi&\sin^2\phi+x&\sin^2\phi-x&-xr\sin\theta&xr\cos\theta\\
\cos\phi\sin\phi&\cos\phi\sin\phi&\sin^2\phi-x&\sin^2\phi+x&xr\sin\theta&-xr\cos\theta\\
xr\cos\theta&-xr\cos\theta&-xr\sin\theta&xr\sin\theta&x(2+r^2)&0\\
xr\sin\theta&-xr\sin\theta&xr\cos\theta&-xr\cos\theta&0&x(2+r^2)
\end{array}
\right).
\eeq
The eigenvalues of this matrix are the masses of the six chiral multiplets.  As in Ref.~\cite{0012250}, one chiral multiplet is massless and is eaten by the Higgs mechanism and one has a mass equal to that of the vector multiplet and is the superpartner of the photon in the massive $\mathcal{N}=1$ vector multiplet. 

The other four masses depend on $x$ and $r$ but not on $\theta$ and $\phi$.  As is required by $\mathcal{N}=1$ supersymmetry, they come in pairs, corresponding to the masses of the complex scalars.  The two distinct masses are
\beq
\frac{m^{\pm\ 2}_\chi}{m^2_\gamma}=x(1+\frac{r^2}{4}\pm \frac{r}{4}\sqrt{8+r^2}). \label{mass2}
\eeq
In particular, at $x=1$ these are the masses (\ref{mass1}) with an effective linear and quadratic term in the the polynomial $F$ reflecting the fact that $G$ may now be nontrivial.  In particular, when $x=1$ the masses $m^-_\chi$ are always less than those of the vector multiplet and so the superconductor is type I and there are no BCS vortices.  From Eq.~(\ref{mass2}) one can see that when $x<1$ than the chiral multiplets are even less massive and so the attractive force between the vortices grows and the superconductor becomes even more type I.

However when $x$ is sufficiently large then
\beq
\frac{1}{x}<1+\frac{r^2}{4}- \frac{r}{4}\sqrt{8+r^2}
\eeq
and the superconductor is type II and admits vortex solutions.   This condition corresponds to $G\p(\phi_0)$ positive and sufficiently large with respect to the adjoint chiral multiplet mass.  In particular, shifting $\phi$ so that $\phi_0=0$, it corresponds to a sufficiently large positive additional $Q\Phi\tilde{Q}$ term in the superpotential.  This term is marginal and so, as we have stressed, the breaking of the $\mathcal{N}=2$ supersymmetry is not soft.  Therefore we conclude that if $\mathcal{N}=2$ supersymmetry is softly broken to $\mathcal{N}=1$ at tree level, and not in a separate sector, then the superconductivity will remain type I and the vortices of the gauge theory will have gradients beyond the range of validity of the Landau-Ginzburg approximation of the BCS superconductor.

\section{Conclusions}

Known vortices in supersymmetric gauge theories exhibit qualitatively different behavior from those of superconductors and Yang-Mills, in particular the former attract and the latter repel.  We have argued that if one attempts to embed a vortex similar to that in a supersymmetric gauge theory into a BCS superconductor, then the nonlocal effects of the effective theory will be important and will not be captured by the SUSY gauge theory.  

This conclusion came with a number of disclaimers.  For one, it is a classical analysis, and one may hope that a repulsive force may be found in the quantum theory.   We have argued that higher order superpotentials corresponding to a soft breaking of $\mathcal{N}=2$ supersymmetry, with our identification of the BCS parameters and the gauge theory parameters, will never lead to a type II superconductor at tree level but may nonetheless lead to repulsive vortices, which may be consistent with lattice results.  Clearly it would be interesting to investigate models with a dynamical breaking of $\mathcal{N}=2$ to $\mathcal{N}=1$ (if they exist) or with extra sectors responsible for the breaking to see if, as in the case of the selectron versus electron mass in the MSSM, such breakings may yield chiral multiplets which are all more massive than the vector multiplets.

We used MQCD to argue that higher order superpotentials do lead to some effects not present in the case of quadratic superpotentials.  However we have argued that in the case of soft $\mathcal{N}=2$ breaking the superconductivity will remain type I or critical, as such superpotentials do not affect the mass terms of the $\mathcal{N}=1$ supermultiplets and so do not change the length scales of the superconductor.  In fact a number of numerical calculations \cite{Walter} confirm that such superpotentials lead to attractive vortices, and that the superpotentials found in Sec.~\ref{gensec} that lead to a hard breaking of the $\mathcal{N}=2$ do indeed lead to repulsive vortices, as we have claimed using an asymptotic argument.


It may be that SQCD fails to describe a pure BCS theory, but describes an impure theory.  For example, attractive vortices in superconductors with two kinds of Cooper pairs have been conjectured in exist in neutron stars in Ref.~\cite{neutron}, and are known to exist in the superconductor $MgB_2$.  In the later case it has even been claimed that, as we have seen is always the case in softly broken $\mathcal{N}=2$ supersymmetric gauge theories, one condensate coherence length is greater than the penetration depth and the other is less \cite{unoemezzo}.  It may be that QCD, having many species of charged particles which may condense, allows attractive vortices.  The additional more massive chiral multiplets do presumably correspond to other species of Cooper pairs which, having shorter coherence lengths, are subdominant in the long distance interactions of vortices.  However these are more massive than the vector multiplets and so they may well dominate the short distance behavior of vortices and therefore lead to type II superconductivity near the vortex core.  This means that the impure BCS theory may be expected to have vortices which attract at large distances due to the light chiral multiplets but nonetheless have a subcritical magnetic field in their cores due to the heavy chiral multiplets.

One may ask whether the difference in vacuum structure is really important.  The importance is that some models of the QCD condensate will not have analogues in a supersymmetric theory with attractive vortices.  For example, it may be that, like type II superconductors in a sufficiently strong background magnetic field, the QCD vacuum has a non-translationally invariant structure  like the Abrikosov lattice whereas all supersymmetric theories that descend from $\mathcal{N}=2$ have translationally-invariant vacua.  In such a scenario the one-loop instabilities of QCD may play the role of the background magnetic flux in the superconductor \cite{NO}.  

\section* {Acknowledgement}

We would like to thank Allan Adams, Walter Vinci and Alexei Yung for enlightening discussions and for a reading of this manuscript, as well as Roberto Auzzi and Sven Bjarke Gudnason.  We thank Giacomo Marmorini for finding an error in an earlier version of the draft.


\end{document}